\documentclass[conference,a4paper]{neutr}
\pdfoutput=1 
\neutrAuthorListTex{Tobias Bauer, Fabian Beer, Daniel Holl, Ardian Imeraj, Konrad Schweiger,\\ Philipp Stangl, Wolfgang Weigl und Christoph P.~Neumann\,\orcidlink{0000-0002-5936-631X}}
\neutrAuthorListBib{Tobias Bauer and Fabian Beer and Daniel Holl and Ardian Imeraj and Konrad Schweiger and Philipp Stangl and Wolfgang Weigl and Christoph P. Neumann}
\neutrTitleTex{{Reddiment: Eine SvelteKit- und ElasticSearch\-/basierte Reddit Sentiment\-/Analyse}}
\neutrTitleBib{{\NeutrTitleTex}}
\neutrDepartment{CyberLytics\-/Lab an der Fakultät Elektrotechnik, Medien und Informatik}
\neutrInstitution{Ostbayerische Technische Hochschule Amberg\-/Weiden}
\neutrAddress{Amberg, Deutschland}
\neutrSeries{Technical Reports}
\neutrYear{2022}
\neutrMonth{7}
\neutrNumber{CL-\NeutrYear-06}
\neutrLang{english}

\filecontentsForceExpansion|[] 

\usepackage{rotating}

\graphicspath{ {./} }

\begin{document}
\selectlanguage{\NeutrLang}

\maketitle

\begin{abstract}
Reddiment ist ein webbasiertes Dashboard, das Sentiment-Analyse von Subreddit-Texten mit Aktienkursen verknüpft. Das System besteht aus Backend, Frontend und verschiedenen Diensten. Das Backend, in Node.js, verwaltet die Daten und kommuniziert mit Crawlern, die Reddit-Kommentare und Aktienmarktdaten sammeln. Sentiment wird mithilfe von Vader und TextBlob analysiert. Das Frontend, basierend auf SvelteKit, bietet Benutzern ein Dashboard zur Visualisierung. Die Verteilung erfolgt über Docker-Container und Docker Compose. Das Projekt bietet Erweiterungsmöglichkeiten, z.B. die Einbindung von Kryptowährungskursen. Reddiment ermöglicht die Analyse von Sentiment und Aktienkursen aus Subreddit-Daten.
\end{abstract}


\section{Einführung und Ziele}

In dem Subreddit \texttt{r/wallstreetbets}, auch bekannt als WallStreetBets oder WSB, wird über Aktien- und Optionshandel spekuliert. Er ist bekannt für seine profane Art und die Vorwürfe, dass Nutzer/innen Wertpapiere manipulieren und volatile Kursbewegungen auslösen. Anhand von Sentiment-Analyse sollen nun Subreddits in Bezug auf Aktienkursverläufe analysiert werden. Dazu soll ein webbasiertes Dashboard entwickelt werden, welches den zeitlichen Verlauf von Sentiment und Erwähnungen benutzerdefinierter Schlüsselwörter in ausgewählten Subreddits dem Aktienverlauf gegenüberstellt.

In den weiteren Abschnitten des Technical Reports wird zuerst auf die Bausteinsicht des Gesamtsystems in Abschnitt~\ref{s:bausteinsicht} eingegangen. Im Nächsten Abschnitt~\ref{s:verteilungssicht} wird die Verteilungssicht der Anwendung beschrieben. In Abschnitt~\ref{s:entwicklungswerkzeuge} werden die angewandten Werkzeuge zur Entwicklung der Anwendung vorgestellt. Abschließend wird kurz das Sicherheitskonzept in Abschnitt~\ref{s:secrets} vorgestellt und ein Fazit in Abschnitt~\ref{s:fazit} gegeben.

\section{Bausteinsicht} \label{s:bausteinsicht}
Diese Sicht zeigt die statische Zerlegung des Systems in Bausteine sowie deren Beziehungen.
Der Quellcode ist als Open-Source veröffentlicht\footnote{\url{https://github.com/cyberlytics/Reddiment}}.

\subsection{Gesamtsystem}
\begin{figure}[ht]
    \centering
    \includegraphics[width=\linewidth]{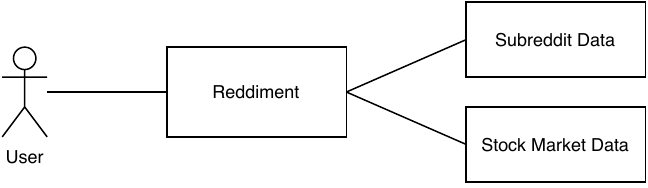}
    \caption{Kontextabgrenzung des Reddiment Gesamtsystems}
    \label{fig:context}
\end{figure}

Reddiment bezieht Daten aus mehreren Quellen und stellt diese dem Benutzer aggregiert bereit. Die folgende Abbildung~\ref{fig:context} zeigt die Interaktionen des Systems mit Fremdsystemen und dem Benutzer. Die Datenquellen für Reddiment sind
\begin{itemize}
  \item Reddit-API für Subreddit-Daten und
  \item Yahoo Finance für Aktienmarkt-Daten.
\end{itemize}

\begin{figure*}[ht]
    \centering
    \includegraphics[width=\linewidth]{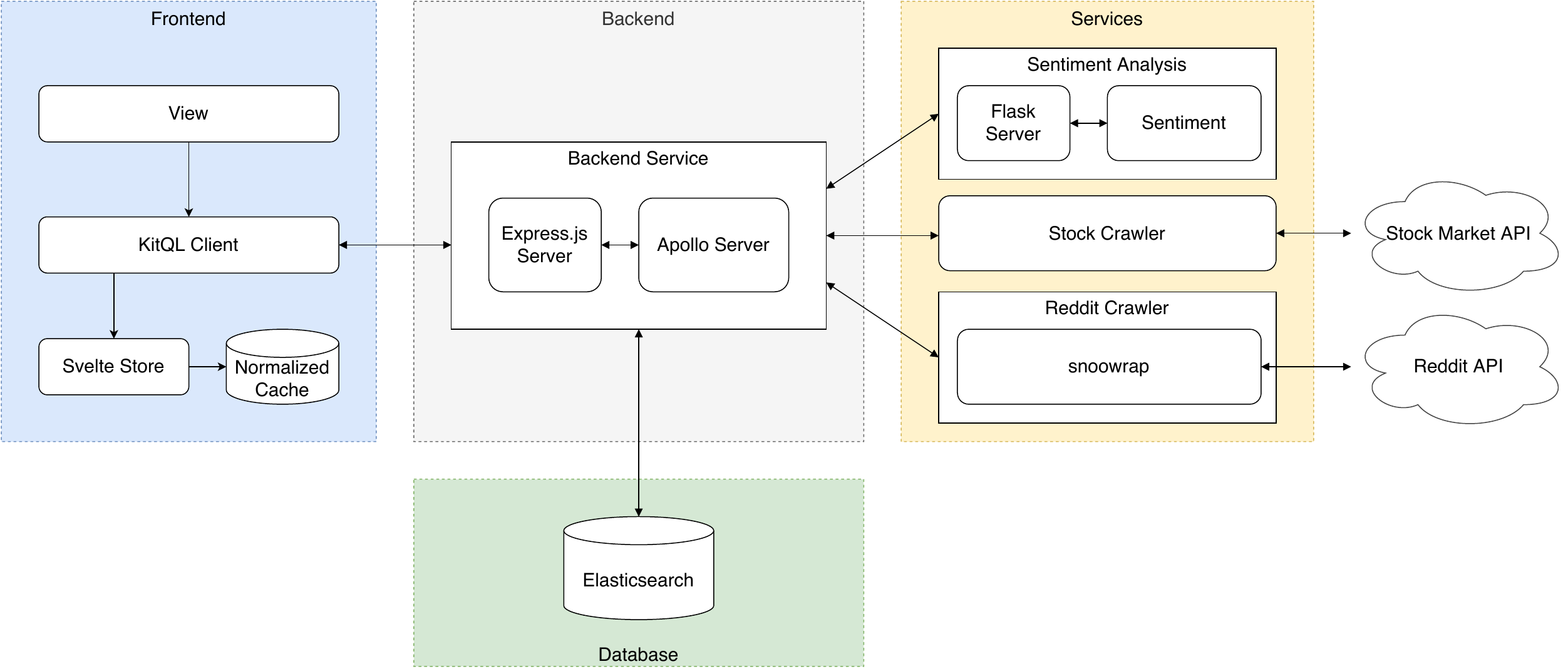}
    \caption{Überblick über die Architektur von Reddiment. Die Architektur besteht aus vier Teilen: Das Frontend bietet dem Nutzer ein graphisches Dashboard zur Visualisierung der Metriken (Abschnitt~\ref{sub:frontend}),  das Backend stellt (Abschnitt~\ref{sub:backend}) die GraphQL-API bereit, die Datenbank (Abschnitt~\ref{sub:database}) speichert persistent Reddit- und Aktienmarkt-Daten, und die Dienste (Abschnitt~\ref{sub:services}) führen die Sentiment-Analyse aus und beschaffen die Rohdaten.}
    \label{fig:architecture}
\end{figure*}

Ein genauerer Überblick über die Architektur von Reddiment findet sich in Abbildung~\ref{fig:architecture}.
Vorarbeiten des CyberLytics-Labs mit in Teilen überlappendem Technologie-Stack waren Covidash \cite{ModA-TR-2021SS-WAE-TeamWeiss-CovidDashboard} und Graphvio \cite{ModA-TR-2022WS-SWT-TeamGruen-Graphvio}.

\subsection{Backend} \label{sub:backend}
Das Backend wurde als ein unter „Node.js“~\cite{node} laufender Server realisiert. Hierfür kam das Web-Framework „Express“~\cite{express} zum Einsatz. Darauf aufbauend wurde eine GraphQL-Schnittstelle~\cite{graphql} mithilfe des Frameworks „Apollo Server“~\cite{apolloserver} implementiert. Die GraphQL-API ist unter der Route \texttt{/graphql} verwendbar.

Das Backend stellt die zentrale Kommunikations- und Verwaltungseinheit für die einzelnen Komponenten des Gesamtsystems Reddiment dar. So liefern die proaktiven Crawler (siehe Abschnitt~\ref{sub:services}) die Rohdaten über eine GraphQL-Mutation an das Backend. Diese Rohdaten werden anschließend weiterverarbeitet und im Falle von Reddit-Kommentaren durch Aufruf des Sentiment-Services um den Sentiment-Wert erweitert. Alle so erhaltenen Daten werden über die Anbindung zur Datenbank (siehe Abschnitt~\ref{sub:database}) persistiert.

Stellt nun das Frontend eine Anfrage an das Backend, werden diese aus der Datenbank geladen und aufbereitet. Die Gruppierung und Aggregation von beispielsweise Reddit-Kommentaren sowie die Interpretation des Sentiment-Werts erfolgt in der Suchanfrage an die Datenbank. Damit wird die rechenaufwendige Aggregation direkt an der Quelle der Daten ausgeführt und gleichzeitig Bandbreite eingespart. Die Konfiguration im Backend hingegen ermöglicht es uns, zukünftig eine feinere oder auch gröbere Zusammenfassung von Sentiment-Daten ohne Datenverlust oder aufwendigen Transformationen vorzunehmen.

Aufgrund des Einsatzes von GraphQL als Schnittstelle können Frontend und ggf. andere (externe) Akteure die benötigten Daten in einer von ihnen gewählten Struktur abfragen. Dies erlaubte eine flexible Entwicklung des Frontends ohne Kommunikations-Overhead. Weiterhin sind server-seitig die sogenannten Resolver modular aufgebaut, sodass verschiedene GraphQL-Felder in getrennten Quellcodedateien behandelt werden können. Alle Quellcodedateien im Zusammenhang mit GraphQL befinden sich im Verzeichnis \texttt{backend/src/graphql}.

Das Backend ist aufgrund der zentralen Position im System Reddiment auf die anderen Bestandteile angewiesen. Um dennoch Unit-Tests für ein isoliertes Backend zu schreiben und auszuführen, können die Datenbank und das Sentiment-Modul dynamisch durch Mocks ersetzt werden. Beim Start des Backends kann über die Umgebungsvariable \texttt{PRODUCTION} festgelegt werden, ob die realen Module (\texttt{true}) oder die Mock-Module (\texttt{false}) verwendet werden. Die entsprechenden Quellcodedateien liegen im Ordner \texttt{backend/src/services}. Da die Crawler proaktiv sind, müssen für diese keine Mocks erstellt werden -- vielmehr simulieren die Unit-Tests die Crawler und prüfen, ob die gewünschten Daten von der GraphQL-Schnittstelle zurückgeliefert werden.

Im Ordner \texttt{backend/src/util} befinden sich verschiedene Quellcodedateien mit Hilfsfunktionen. Gestartet wird das Backend durch Ausführen der Datei \texttt{backend/server.ts}, die wiederum alle benötigten Module lädt und den Apollo Server mitsamt der GraphQL-Schnittstelle startet.

\subsection{Datenbank} \label{sub:database}
In der NoSQL-Datenbank „Elasticsearch“~\cite{elasticsearch} werden Daten in Form von Dokumenten gespeichert, wobei zusätzliche Indizes einen konsistenten Zugriff auf die enthaltenen semi-strukturierte Daten (vgl.\ \cite{NeFL10oxdbs}) erlauben.
Für dieses Projekt werden Daten, welche an das Backend gesendet werden, als Dokumente persistent in der Datenbank gespeichert. Die Reddit-Daten bestehen aus den neun Feldern in Tabelle~\ref{tab:reddit}.
\begin{table}[h!]
\begin{center}
\caption{Reddit-Daten}
\label{tab:reddit}
\begin{tabular}{|l|l|}
\textbf{Feld} & \textbf{Beschreibung}\\
\hline
\texttt{subreddit}         & Name des Subreddits\\
\texttt{text}             & Kommentar als String \\
\texttt{timestamp}         & Zeitstempel des Kommentars\\
\texttt{commentId}         & Reddit-Kommentar-ID\\
\texttt{userId}         & Reddit-User-ID\\
\texttt{articleId}         & Reddit-Post-ID\\
\texttt{upvotes}         & Anzahl Postiv-Bewertung in Reddit\\
\texttt{downvotes}         & Anzahl Negativ-Berwertung in Reddit \\
\texttt{sentiment}         & Ausgabe der Sentiment-Analyse\\
\end{tabular}
\end{center}
\end{table}

Die Aktienmarkt-Daten setzen sich aus den acht Feldern in Tabelle~\ref{tab:stock} zusammen.

\begin{table}[h!]
\begin{center}
\caption{Aktienmarkt-Daten}
\label{tab:stock}
\begin{tabular}{|l|l|}
\textbf{Feld} & \textbf{Beschreibung}\\
\hline
\texttt{stock}        & Name der Aktie\\
\texttt{timestamp}    & Zeitstempel der Werte\\
\texttt{open}        & Wert bei Tagesbeginn\\
\texttt{high}        & Maximaler Tages-Wert\\
\texttt{low}        & Minimaler Tages-Wert\\
\texttt{close}        & Wert bei Tagesabschluss\\
\texttt{volume}        & Marktvolumen der gehandelten Aktie\\
\end{tabular}
\end{center}
\end{table}

Um die Daten in der Datenbank zu organisieren, arbeitet Elasticsearch mit Indizes. Dies bedeutet, dass ähnlich strukturierte  Dokumente unter demselben Index gespeichert werden. Pro Subreddit gibt es einen Index, dessen Name mit \texttt{r\_} präfigiert ist, und pro Aktie gibt es einen Index mit Präfix \texttt{f\_}. Um einen zeitlichen Verlauf einer Aktie oder des Sentiments zu erzeugen, müssen alle Dokumente eines Index abgerufen werden. Dieses Vorgehen ermöglicht es, dass bei einer Suchanfrage nicht alle Dokumente aller Indizes durchsucht werden müssen, sondern eine Vorauswahl auf eine Untermenge getroffen werden kann.

Elasticsearch verwendet eine global eindeutige ID für jedes Dokument. Bei Dokumenten, welche Reddit-Daten enthalten, wird die ID des Dokumentes gleich der eindeutigen \texttt{commentId} gesetzt. Den Dokumenten mit Aktien-Daten wird eine ID nach folgendem Schema zugewiesen: \texttt{Aktienname\_Zeitstempel}. Dokumente können auch anhand ihrer ID gesucht und ausgegeben werden. Dies wird beispielsweise bei der Aktualisierung von Dokumenten angewendet.

Im Backend wird das Paket „Elasticsearch Node.js client“~\cite{elasticnode} verwendet. Das Paket enthält einen Client, der eine Verbindung zur Datenbank herstellt. In der TypeScript-Datei \texttt{backend/src/services/database.ts} befindet sich der Quellcode zur Kommunikation mit der Datenbank.

\subsection{Dienste} \label{sub:services}

Dieser Abschnitt beschreibt die Dienste (engl. Services) des Gesamtsystems Reddiment. Es gibt drei Dienste: Sentiment, Reddit-Crawler und Stock-Market-Crawler.

\subsubsection{Sentiment}

Dieser Dienst ermittelt für einen Text das Sentiment. Es wurde eine REST-API \cite{restful} mit dem Endpunkt \texttt{/sentiment} implementiert. An diesen Endpunkt kann mittels einer POST-Anfrage ein Text übertragen werden. Der Dienst ermittelt anschließend die Stimmungslage des Textes und gibt das Ergebnis im JSON-Format zurück. Die Ermittelung des Sentiments erfolgt durch zwei regelbasierte Verfahren, um eine höhere Aussagekraft zu haben. Das erste regelbasiertes Verfahren ist „vader“~\cite{vader} (Valence Aware Dictionary and Sentiment Reasoner) der Python-Bibliothek \textit{nltk}.  Das zweite Verfahren ist „TextBlob“~\cite{textblob}. Sind sich die beiden Verfahren in der Auswertung einig, wird der Sentiment-Wert von vader zurückgegeben. Andernfalls wird 0 (neutral) zurückgegeben.

\subsubsection{Reddit-Crawler}

Der Reddit-Crawler verwendet das Paket „snoowrap“~\cite{snoowrap}, das JavaScript-Funktionen für den Zugriff auf die Reddit-API bereitstellt. „snoowrap“ unterliegt den API-Regeln von Reddit, worin u.a. das Rate-Limit bei 60 Anfragen pro Minute liegt. Um die Reddit-API nutzen zu können, muss ein API-Key über das Benutzerkonto\footnote{\url{https://www.reddit.com/prefs/apps}} angefordert werden. Ein API-Key besteht aus den zwei Teilen: einer \textit{Client-ID} und einem \textit{Client-Secret}.

Mit dem API-Key und den Anmeldedaten für das zugehörige Reddit-Konto kann ein Objekt der Klasse \texttt{Snoowrap} erstellt werden. Mit diesem Objekt können Reddit-API-Aufrufe durchgeführt werden.

Der Reddit-Crawler stellt in einem zeitlichen Intervall eine Anfrage an das Backend und erhält eine Liste mit Subreddit-Namen. Kommentare dieser Subreddits sollen von Reddit abgefragt und an das Backend gesendet werden.
Das Sammeln der Kommentare erfolgt zyklisch, um das Rate-Limit nicht zu überschreiten. Pro Sammelzyklus werden die Kommentare an das Backend übermittelt.

\subsubsection{Stock-Market-Crawler}

Um den Verlauf des Sentiments in der Vergangenheit auch mit der tatsächlichen Marktlage vergleichen zu können,
gibt es einen proaktiven Stock-Market-Crawler. Dieser stellt, wie der Reddit-Crawler, in einem zeitlichen Intervall eine Anfrage an das Backend und erhält eine Liste mit  \textit{Tickersymbolen}(Aktienkürzel). Der zugehörige Aktienkurs wird von \textit{Yahoo-Finance} abgefragt und zurückgegeben.

\subsection{Frontend} \label{sub:frontend}
Dieser Abschnitt beschreibt die client-seitige Frontend-Architektur. Das Frontend wird unter Zuhilfenahme des Frontend-Frameworks „SvelteKit“~\cite{sveltekit} realisiert. Das Frontend besteht aus zwei Bausteinen: 1) den Routen zur Navigation, und 2) der \textit{Library} für Komponenten und weitere Module.

\subsubsection{Routen}

Die Routen zur Navigation sind durch Svelte-Kompontenen in \texttt{src/routes} festgelegt. Es gibt Elemente, die auf jeder Seite sichtbar sind, z. B. die Navigationsleiste oder eine Fußzeile. Anstatt diese für jede Seite neu zu definieren,  wird eine Layout-Komponente namens \texttt{src/routes/\_\_layout.svelte} verwendet.

\subsubsection{Library}

Die \textit{Library} (\texttt{src/lib}) ist eine Sammlung von Frontend-Komponenten,  mit denen die eigentliche Anzeige im Webbrowser realisiert wird. Die Komponenten behandeln alle Eingaben und kommunizieren bei Bedarf mit dem Backend über die GraphQL API. Für die Kommunikation mit der GraphQL API wird im Frontend „KitQL“~\cite{kitql} verwendet, das einen client-seitigen GraphQL Client bereitstellt.

\section{Verteilungssicht} \label{s:verteilungssicht}
Zentraler Bestandteil der Verteilungsstruktur sind Docker-Container~\cite{docker}. Jeder Baustein von Reddiment ist für sich isoliert in einem eigenen Docker-Container untergebracht.
Damit das gesamte System mit allen verteilten Komponenten in der korrekten Reihenfolge gestartet wird, werden die Docker-Container mit Docker Compose orchestriert.
Docker Compose übernimmt die Netzwerkkonfiguration und die Vergabe von Host-Namen an die jeweiligen Docker-Container.
Darüber hinaus können Umgebungsvariablen verwaltet werden. Sensible Daten, wie Zugangspasswörter, werden durch spezielle Mechanismen (siehe Abschnitt~\ref{s:secrets}) sicher zur Verfügung gestellt.

\section{Entwicklungswerkzeuge} \label{s:entwicklungswerkzeuge}

In diesem Abschnitt wird auf die verwendeten Entwicklungswerkzeuge genauer eingegangen.

\subsection{Paketverwaltung}

Die Verwaltung der Abhängigkeiten erfolgt mit „npm“ \cite{npm} für auf „Node.js“ basierende Bausteine. Für den Sentiment-Baustein wird „Pipenv“~\cite{pipenv} verwendet.

\subsection{Linting}

Im Frontend wird „eslint“~\cite{eslint} in Verbindung mit „prettier“~\cite{prettier} verwendet,um die Einhaltung der Codierrichtlinien zu gewährleisten. Die Konfigurationen sind jeweils in den Dateien \texttt{eslintrc.js} und \texttt{prettierrc.js} hinterlegt.

\subsection{Build-Tools}

\subsubsection{Backend}

Sowohl das Backend als auch die Crawler werden mit dem TypeScript Compiler „tsc“~\cite{typescript} in ein für „Node.js“ ausführbares Format überführt. Die Konfiguration befindet sich dabei in der Datei \texttt{tsconfig.json}.

\subsubsection{Frontend}
Im Frontend ist Vite dafür zuständig, die Anwendung aus dem Quellcode zu erstellen. Dabei gibt es zwei Varianten: Für Entwicklungszwecke wird ein Vite-Dev-Server (mit Reload-Funktionalität) zum Bereitstellen der Anwendung verwendet. Für den Produktiveinsatz werden nur die benötigten Zieldateien unter Verwendung des \texttt{Static adapter} erstellt, die dann mit einer beliebigen Server-Software ausgeliefert werden können.

\subsection{Unit-Tests}

Im Backend werden Unit-Tests mit „mocha“~\cite{mochajs} ausgeführt. Dabei wird „istanbul“ \cite{istanbuljs} für die Erzeugung der Test-Abdeckung verwendet.

Im Frontend wird „Vitest“~\cite{vitest} verwendet. Für die Frontend-Komponenten wird zusätzlich die „svelte-testing-library“~ \cite{stl} eingesetzt. Diese ermöglicht es, die Komponenten zu \textit{rendern}, um Details über die verschiedenen Elemente innerhalb der Komponente zu erhalten.

\section{Secrets-Verwaltung} \label{s:secrets}

Die Anforderung, sensible Daten jeder Anwendung individuell und zusätzlich einfach konfigurierbar zur Verfügung zu stellen, lässt schnell auf den Einsatz von Umgebungsvariablen schließen. Ein solches Vorgehen birgt jedoch einige sicherheitsrelevante Schwachstellen. Aus diesem Grund werden die komponentenspezifischen Zugangsdaten mittels dem Schlüsselwort \texttt{secrets} in der Docker-Compose-Konfigurationsdatei den Docker-Containern zur Verfügung gestellt. Dabei werden im Gegensatz zu Umgebungsvariablen die Passwörter in einer Text-Datei gespeichert und in dem jeweiligen Docker-Container eingebunden.

Ein weiteres zu beachtendes Kriterium in diesem Sicherheitskonzept ist der Zugriff auf die jeweiligen API der einzelnen Komponenten. Beispielsweise sollen nur die Crawler die GraphQL-Mutations des Backends verwenden dürfen. Dafür wurde für jede Anwendung ein individueller „Access-Key“ erstellt. Der Zugriff wird durch Prüfen des Schlüssels bei jeder Anfrage gewährt bzw. unterbunden. Unberechtigte Dritte können dadurch die API nicht verwenden und ein Missbrauch wird dadurch erschwert. Aufgrund zeitlicher Einschränkungen konnte das Sicherheitskonzept nicht vollständig implementiert werden.

\section{Fazit und Ausblick} \label{s:fazit}

Reddiment ist ein webbasiertes Dashboard, welches den zeitlichen Verlauf von Sentiment und Erwähnungen bestimmter Schlüsselwörter in ausgewählten Subreddits dem Aktienverlauf gegenüberstellt. Durch die Auswahl eines Subreddits und eines oder mehrere Schlüsselwörter werden die Diagramme automatisiert mit den vorhandenen Daten befüllt.
Der Reddit-Crawler, zur kontinuierlichen Beschaffung von Reddit-Daten, hat bei der Umsetzung die meiste Zeit in Anspruch genommen. Gründe dafür waren zum einen fehlende Vorkenntnisse in der Programmiersprache Typescript, aufgrund heterogener Vorkenntnisse der Team-Mitglieder, zum anderen gab es Probleme mit der Dokumentation des verwendeten Moduls „snoowrap“.  

Aufgrund der strikten Trennung von Front- und Backend besteht die Möglichkeit, verwendete APIs zu ersetzen oder Weitere einzubinden. So ist es beispielsweise denkbar, neben den Aktienmarkt-Daten, Kursdaten für Kryptowährungen einzubinden. Dadurch könnte ein Vergleich von Sentiment und Erwähnungen gegenüber dem Kursverlauf einer Kryptowährung wie z.B. Bitcoin durchgeführt werden.


\printbibliography[notcategory=selfref]

\newpage

\begin{sidewaysfigure*}
  \includegraphics[page=1, width=\columnwidth]{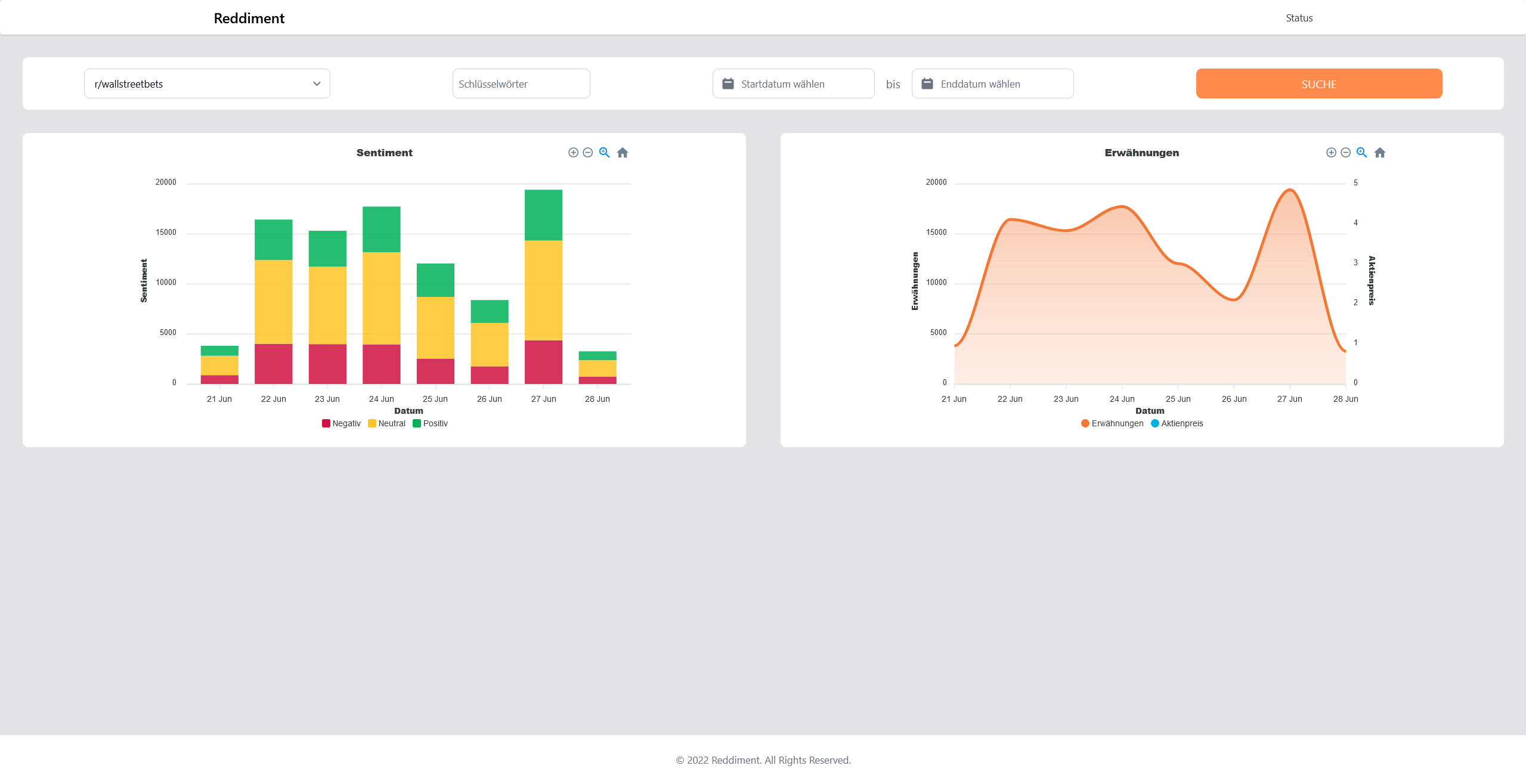}
  \caption{Das Reddiment Dashboard. Die abgebildete Suchanfrage enthält keine Schlüsselwörter, daher werden Sentiment und Erwähnungen aller Einträge in \texttt{r/wallstreetbets} angezeigt.}
  \label{fig:extended_layers}
  \end{sidewaysfigure*}

\end{document}